\documentclass[aps,prl,twocolumn,superscriptaddress,nofootinbib]{revtex4-2}
\usepackage{amsmath,amssymb,graphicx,booktabs}
\usepackage[hidelinks]{hyperref}
\usepackage{soul}
\usepackage[normalem]{ulem}
\usepackage{xcolor}
\usepackage{enumitem}
\usepackage{multirow}

\newcommand{\Om}{\Omega_m}
\newcommand{\fs}{f\sigma_8}
\newcommand{\zpk}{z_{\rm pk}}
\newcommand{\Opk}{\Omega_{m(\rm pk)}}
\newcommand{\fpk}{f_{\rm pk}}
\newcommand{\xpk}{x_{\rm pk}}
\newcommand{\RL}{R_\Lambda}

\newcommand{\add}[1]{{\color{blue}#1}}

\newcommand{\del}[1]{%
  \ifmmode
    \text{\color{red}\sout{$#1$}}%
  \else
    {\color{red}\sout{#1}}%
  \fi
}
\definecolor{pink}{RGB}{185,25,115}

\definecolor{reviewpink}{RGB}{185,25,115}
\definecolor{reviewgreen}{RGB}{0,100,50}
\newcommand{\newdel}[1]{%
  \ifmmode
    \text{\color{reviewpink}\sout{$#1$}}%
  \else
    {\color{reviewpink}\sout{#1}}%
  \fi
}

\begin{document}

\title{A Universal Ratio Linking Cosmic Expansion and the Growth of Structure}
\author{Anjan A Sen}
\email{aasen@jmi.ac.in}
\affiliation{%
Centre for Theoretical Physics, Jamia Millia Islamia, New Delhi 110025, India
}

\author{Purba Mukherjee}
\email{purba@kasi.re.kr}
\affiliation{%
Korea Astronomy and Space Science Institute (KASI), Daejeon 34055, Republic of Korea%
}

\date{\today}

\begin{abstract}
We identify a universal relation between the onset of cosmic
acceleration ($z_t$) and the turnover
($z_{\rm pk}$) of the linear growth
observable $\fs$. In a spatially flat $\Lambda$CDM Universe,
these two redshifts obey
$R\equiv(1+z_t)/(1+\zpk)=1.1596$, independent of the present
matter fraction and the overall expansion and growth
normalisations. This provides a new consistency test for $\Lambda$CDM,
linking both background and growth dynamics.
We further establish
the conditions under which this extends to broader classes of
dark energy and modified gravity theories. 
\end{abstract}

\maketitle

\noindent
\textit{\underline{Introduction}:}
The physical origin of the late-time accelerated expansion
remains an open question in cosmology%
~\cite{Riess1998,Perlmutter1999}.
Although the $\Lambda$CDM model gives the simplest
description, recent DESI observations, combined with cosmic
microwave background (CMB) and Type Ia supernova (SN-Ia)
data, have renewed interest in evolving dark energy%
~\cite{DESIDR2}. On the other hand, the
degeneracy between dark energy and modified gravity in the
background expansion motivates joint tests of cosmic
expansion and structure growth%
~\cite{Linder2005,Guzzo2008}.

The onset of cosmic acceleration and the turnover of linear structure growth represent two distinct 
epochs of the cosmological evolution determined by evolving energy density components. While the former 
is determined by the background expansion, 
the latter exhibits the competition between gravitational clustering and Hubble friction. This raises a natural
question: does the separation between these two epochs depend on the present matter abundance, or 
does it represent a more fundamental relation between expansion and growth? 
In this {\it letter}, we first establish such a relation in a spatially flat $\Lambda$CDM model and 
then investigate the conditions under which it extends to broader families of dark energy and modified gravity (MG) models.
\smallskip

\begin{figure*}[t]
\centering
\includegraphics[width=0.45\textwidth]{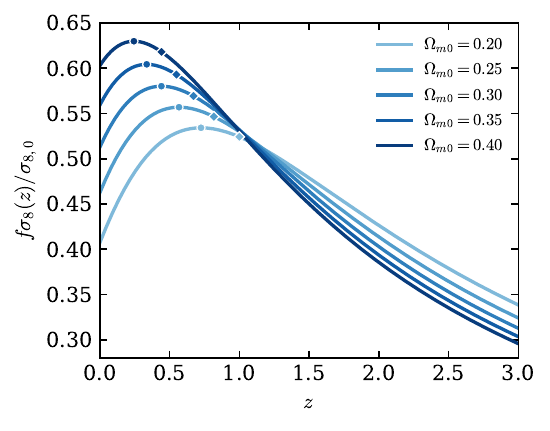}\hfill
\includegraphics[width=0.45\textwidth]{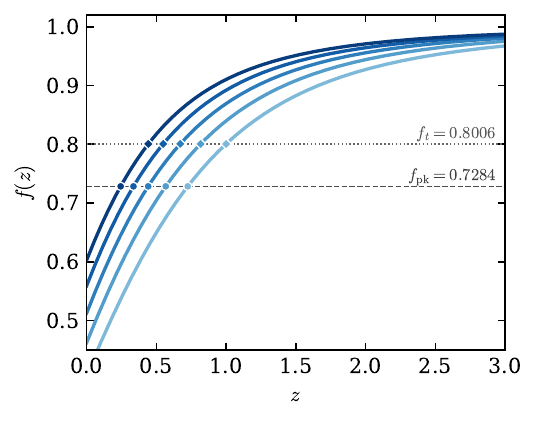}
\caption{Flat $\Lambda$CDM, $\Omega_{m0}=0.20$--$0.40$
(light to dark). Circles: maximum of $\fs$; diamonds: onset of acceleration.
Left Figure: $\fs/\sigma_{8,0}$ vs $z$. Right Figure: Growth rate $f$ vs $z$: $f=0.7284$ at every $\fs$ maximum,
$0.8006$ at every onset of acceleration.}
\label{fig:lcdm}
\end{figure*}
\noindent
\textit{\underline{Theoretical framework}:}
We consider a spatially flat FLRW universe with scale factor
$a=(1+z)^{-1}$ and background expansion rate $H=\dot a/a$.
Defining $N\equiv\ln a$, the deceleration parameter is
$q\equiv-\ddot a/(aH^2)=-1-{\rm d}\ln H/{\rm d}N$.
The matter density parameter is
$\Om\equiv8\pi G\rho_m/(3H^2)$, and the total effective
equation of state (EoS),
$w_T\equiv p_{\rm tot}/\rho_{\rm tot}$, satisfies
$q=\tfrac12(1+3w_T)$. In MG, $\rho_{\rm tot}$ and $p_{\rm tot}$ include
the effective contributions obtained by writing the
background equations in Einstein form. 
Throughout, primes denote ${\rm d}/{\rm d}N$, and
subscript $0$ denotes present-day values ($a=1$).
We work in the late-time approximation,
neglecting radiation.

Matter perturbations at linear order grow as
$\delta\propto D(a)$, with growth factor $f\equiv D'/D$.
The quantity $\sigma_8$ is the root mean squared (rms) linear
density fluctuation smoothed within spheres of comoving
radius $8h^{-1}\,\mathrm{Mpc}$.
In MG scenarios, modifications to the gravitational
clustering strength are encoded through
$\mu\equiv G_{\rm eff}/G$, where $G_{\rm eff}$ is the
effective gravitational strength and $\mu=1$ for
pressureless matter and smooth dark energy in GR.%
\footnote{In general, $\mu=\mu(k,a)$, where $k$ is the
comoving wavenumber. We consider the linear regime in
which its scale dependence can be neglected over the
relevant range of $k$, so that $\mu(k,a)\simeq\mu(a)$
and a scale-independent growth factor $D(a)$ applies%
~\cite{BelliniSawicki2014}.}

Type Ia supernovae~\cite{Riess1998,Perlmutter1999} and
baryon acoustic oscillations (BAO)~\cite{Eisenstein2005}
provide complementary constraints on the background
expansion, including the onset of late-time acceleration.
In contrast, redshift-space distortions (RSD)
probe linear structure growth through $\fs$~\cite{Kaiser1987}.
Combining these observables allows a joint test of the
background expansion and the growth of large-scale
structure.
\smallskip

\noindent
\textit{\underline{Two characteristic cosmic epochs}:}
\begin{enumerate}[left=0pt]
\item[(a)]
The onset of acceleration, $z_t$, defined by
$q(z_t)=0$, is determined from background observables. 
This epoch depends solely on the shape of the expansion 
history and is unchanged by an overall rescaling of 
$H(z)$ through $H_0$ calibration or the BAO sound horizon 
$r_d$.

\item[(b)]
The growth of linear matter perturbations is governed by
\begin{equation}
\begin{aligned}
D''+(1-q)D' &= \tfrac32\mu\Om D,\\
f'+f^2+(1-q)f &= \tfrac32\mu\Om.
\end{aligned}
\label{eq:growth}
\end{equation}
Because
$\fs(z)=\sigma_{8,0}fD/D_0=\sigma_{8,0}D'/D_0$,
the observable $\fs$ is stationary when $D''=0$,
i.e.\ where $f'+f^2=0$.
The turnover corresponds to a change from $D''>0$ to
$D''<0$ as $N$ increases, so that $\fs$ reaches a maximum.
Substituting this condition into Eq.~\eqref{eq:growth},
one obtains an exact relation at the maximum, within the adopted growth description:
\begin{equation}
\fpk=\frac{3\mu\Om}{2(1-q)}
=\frac{3\mu\Om}{1-3w_T}
\quad\text{at }\zpk.
\label{eq:id}
\end{equation}
The peak redshift $\zpk$ defines our second characteristic
epoch, present in the growth history.
It is independent of the overall growth normalisation
$\sigma_{8,0}$, and is unchanged by any constant multiplicative
rescaling of the inferred $\fs$.
\end{enumerate}

\noindent
\textit{\underline{Universal diagnostic for $\Lambda$CDM}:}
We now evaluate the relation between the two epochs in
flat $\Lambda$CDM, for which $\mu=1$.
We define the density ratio
$x\equiv\rho_\Lambda/\rho_m=Ka^3$, with
$K\equiv\Omega_{m0}^{-1}-1$.
Relevant dimensionless background quantities depend on
$x$ alone; in particular,
$\Om=(1+x)^{-1}$ and
$q=(1-2x)/[2(1+x)]$, while $\Omega_{m0}$ appears only
in the constant $K$.
With ${\rm d}/{\rm d}N=3x\,{\rm d}/{\rm d}x$,
both equations in Eq.~\eqref{eq:growth} can be expressed
in terms of $x$, eliminating their explicit dependence on $\Omega_{m0}$.
The resulting growing solution, with $f\to1$ deep in
the matter era, is therefore a function only of $x$%
~\cite{Heath1977,SilveiraWaga1994}:
\begin{equation}
f(x)=1-\frac{6}{11}x
\frac{{}_2F_1\left(\tfrac43,2;\tfrac{17}{6};-x\right)}
{{}_2F_1\left(\tfrac13,1;\tfrac{11}{6};-x\right)}.
\label{eq:grow}
\end{equation}
Here ${}_2F_1$ is the hypergeometric function.
The turnover condition for
$\fs$, given by Eq.~\eqref{eq:id}, reduces to
$\fpk=3/(1+4\xpk)$.
Equating this expression to the growing-mode solution
in Eq.~\eqref{eq:grow}, evaluated at $\xpk$, gives a
transcendental equation for $\xpk$.
Its solution yields the following universal values for
flat $\Lambda$CDM:
\begin{equation}
\xpk=0.7797,\quad
\Opk=0.5619,\quad
\fpk=0.7284.
\label{eq:fpk}
\end{equation}
The peak redshift,
$\zpk=(K/\xpk)^{1/3}-1$, depends on $\Omega_{m0}$; 
$\zpk$ moves from $0.72$ to $0.24$ as
$\Omega_{m0}$ varies from $0.20$ to $0.40$
[Fig.~\ref{fig:lcdm}(a,b)].
Although $\fpk$ is universal in flat $\Lambda$CDM,
its direct measurement is complicated by degeneracies
with galaxy bias and the fluctuation amplitude%
\footnote{RSD measurements primarily determine
$b\sigma_8$ and $f\sigma_8$ rather than $f$ alone,
where $b$ is the galaxy bias. Thus $\fpk$ cannot be
extracted without $b$ and $\sigma_{8,0}$}
The peak redshift $\zpk$, however, can be inferred from
the shape of $\fs(z)$ without fixing its overall
normalisation, making it a useful observable.

At $z=z_t$, $q=0$ implies $x_t=1/2$.
Since $x=Ka^3$ at every epoch, $K$ cancels out in the
ratio of scale factors, giving
\begin{equation}
R\equiv\frac{1+z_t}{1+\zpk}
=\left(\frac{\xpk}{x_t}\right)^{1/3}
=1.1596.
\label{eq:R}
\end{equation}
For example, in $\Lambda$CDM,
$z_t=0.671$ and $\zpk=0.441$ for $\Omega_{m0}=0.3$,
and $z_t=0.817$, $\zpk=0.567$ for $\Omega_{m0}=0.25$:
both epochs move with $\Omega_{m0}$, but
their ratio is fixed at $\RL=1.1596$.

Several existing diagnostics comparing growth
with background expansion involve cosmological parameters.
The growth index $\gamma$, defined by
$f=\Om^\gamma$~\cite{Linder2005}, involves $\Om(z)$,
while the standard GR prediction for the $E_G$ statistic~\cite{Zhang2007}
depends on $\Omega_{m0}$ and $f$.
Other consistency tests compare parameters
(e.g.\ $\Omega_{m0}$) fitted separately to the background
and growth histories~\cite{RuizHuterer2015}.
The peak redshift of $\fs$ has also been proposed as a
$\Lambda$CDM test~\cite{Zheng2022}, but requires
$\Omega_{m0}$ for its theoretical prediction.

The ratio $R$ is, to our knowledge, the first diagnostic
combining the acceleration transition and the growth
turnover to yield a fixed prediction for flat
$\Lambda$CDM, $\RL=1.1596$, independent of $\Omega_{m0}$,
the Hubble constant $H_0$, the sound horizon $r_d$,
and the growth normalisation $\sigma_{8,0}$.
An overall rescaling of $H(z)$ leaves $z_t$ unchanged,
while a constant multiplicative rescaling of $\fs$
leaves $\zpk$ unchanged.
It rests on four assumptions: spatial flatness,
a $\Lambda$CDM background, growth governed by GR,
and the growing-mode condition $f\to1$
deep in the matter era.
A statistically significant, model-independent
estimate of $R\neq\RL$
would therefore rule out flat $\Lambda$CDM with
standard growing-mode evolution, without external
constraints on $\Omega_{m0}$, $H_0$, or $\sigma_{8,0}$.
These invariances remove overall calibration
factors, but not reconstruction uncertainty or
redshift-dependent systematics.

Beyond its role as a consistency test, $\ln R$ has a
direct physical interpretation.
Equation~\eqref{eq:growth} can be written as
$D''/D=\tfrac32\mu\Om-(1-q)f$, which sets the attractive
pull of matter against Hubble friction in the growth
of the matter density contrast per logarithmic scale
factor.
$D'$ increases while this \emph{net pull} is positive,
and $\fs$ peaks when it vanishes
and changes sign.
At the onset of acceleration, the net pull is still
positive ($0.1994$ in $\Lambda$CDM, compared to $1$
in the matter era).
A further $\Delta N=\ln R=0.1481$ $e$-folds of expansion
are required for the damping term to balance the
gravitational source, during which $\Om$ decreases
from $2/3$ to $\Opk=0.5619$.
Thus, $\ln R$ is the number of $e$-folds for which the
net pull outlasts the onset of acceleration.
$R>\RL$ corresponds to a longer separation
after acceleration begins, while $1<R<\RL$
corresponds to a shorter separation.
For $R<1$, the net pull vanishes before acceleration
begins, so the growth turnover precedes the acceleration
transition, as in the DGP example (see Table. \ref{tab:models}).
\smallskip

\noindent
\textit{\underline{The universality class}:}
Building on the flat $\Lambda$CDM result, we now identify
a class of dark energy and modified gravity models for
which $R$ remains independent of $\Omega_{m0}$ and the
overall expansion and growth normalisations, and depends
only on the fixed dimensionless model parameters and
early-time initial conditions. 
The essential property is that varying $\Omega_{m0}$
shifts the dimensionless evolution in $N$ without
changing the trajectory.

\noindent
{\it In flat $\Lambda$CDM, relevant dimensionless
quantities $\Om$, $q$ and $f$ depend only on
$x=Ke^{3N}$. So any epoch defined through a physical
condition on them occurs at a fixed $x$.
The two characteristic epochs are $x_t=\tfrac12$
and $\xpk=0.7797$, respectively.
Changing $\Omega_{m0}$ replaces $K$ by $\widetilde K$,
which shifts each epoch according to
$N\to N-\tfrac13\ln(\widetilde K/K)$.
Since both epochs shift by the same amount for different
values of $\Omega_{m0}$, their separation in $e$-folds,
$\ln R=\tfrac13\ln(\xpk/x_t)$, is independent of
$\Omega_{m0}$ and therefore universal.}

For \emph{any} spatially flat late-time Universe with
separately conserved pressureless matter and the adopted
linear-growth description, the equations governing the
background and growth dynamics are
\begin{equation}
\Om'=3w_T\Om,\qquad
f'=\tfrac32\mu\Om-f^2-\tfrac12(1-3w_T)f.
\label{eq:gen}
\end{equation}
The two epochs satisfy
$w_T=-\tfrac13$
at $z_t$
and $f(1-3w_T)=3\mu\Om$
at $\zpk$.
As $w_T<0$ between these epochs, 
$\Om$ decreases monotonically and serves as a clock.
This allows us to write
\begin{equation}
\ln R=
\int_{\Opk}^{\Omega_{m,t}}
\frac{{\rm d}\Om}{3|w_T(\Om)|\Om},
\label{eq:Rint}
\end{equation}
where $\Omega_{m,t}\equiv\Om(z_t)$.
The integral is positive when the growth
turnover follows acceleration and negative when
it precedes it.
In $\Lambda$CDM, $w_T=\Om-1$, and
Eq.~\eqref{eq:Rint} recovers Eq.~\eqref{eq:R}.

The independence of $R$ from $\Omega_{m0}$ follows
from a scaling property shared by a broad class of
cosmological models.
Suppose the dark energy or modified gravity sector
depends on a \emph{single independent} 
density scale $\rho_*$.
This can be the vacuum energy $\rho_\Lambda$,
the height $V_0$ of a scalar-field potential,
or the crossover scale $r_c$ of a braneworld model
through $\rho_*\propto1/(Gr_c^2)$, 
with all remaining parameters and the
early-time state
(e.g.\ a field frozen in the matter era or on its
tracker attractor) fixed by dimensionless numbers.
For models whose background and growth
equations are invariant under the rescaling 
\begin{equation}
\rho_m\to c\rho_m,\quad
\rho_*\to c\rho_*,\quad
t\to t/\sqrt{c}
\quad(H\to\sqrt{c}H),
\label{eq:scaling}
\end{equation}
with all additional dynamical variables
transformed consistently, the overall dimensional
normalisation does not affect the dimensionless evolution.
For fixed dimensionless theory parameters and
prescribed early-time conditions selecting a common
cosmological trajectory, the background and growth
quantities $\Om$, $w_T$, $\mu$, and $f$ can be expressed
as unique functions of $y\equiv\rho_m/\rho_*$.
Since conservation of pressureless matter gives
$y=y_0e^{-3N}$, both characteristic epochs occur at
fixed values of $y$, independent of $y_0$.
Consequently,
\begin{equation}
R=e^{N_{\rm pk}-N_t}
=\left(\frac{y_t}{y_{\rm pk}}\right)^{1/3},
\label{eq:Rgeneral}
\end{equation}
independently of $y_0$.

This argument requires no analytic solution and applies
to fluid, scalar-field, and modified gravity models
whose background and growth dynamics possess the stated
scaling and common-trajectory properties.
This universality need not hold for parametrisations
anchored today, such as the Chevallier--Polarski--Linder
(CPL) form $w=w_0+w_a(1-a)$%
~\cite{CPL2001,Linder2003},
for which varying $\Omega_{m0}$ at fixed
$(w_0,w_a)$ changes the dimensionless trajectory. 
Spatial curvature introduces
an additional background contribution
whose relative normalisation can likewise alter $R$. 
However, for $\Omega_{k0}$ within present
observational bounds, the resulting shift in $R$
is far below one per cent for the $\Lambda$CDM model.
Similarly, massive neutrinos introduce an additional
scale and mildly scale-dependent growth; for
$\Sigma m_\nu=0.06$\,eV, the resulting shift in $R$
is order $10^{-3}$, well below projected sensitivities.
\medskip

\begin{table}[t]
\caption{Model predictions:
$\Delta R\equiv R/\RL-1$. For universal families,
$R$ is independent of $\Omega_{m0}$; extended results
together with details are given in the Supplemental Material.
All the numbers mentioned related to $R$ 
are for $\Omega_{m0}=0.2,0.3,0.4$. $w_{\rm eff,0}$ is the
present effective DE EoS with quoted value for $\Omega_{m0}=0.3$.}
\begin{ruledtabular}
\begin{tabular}{lccc}
model & $R$ & $\Delta R$ & $w_{\rm eff,0}$\\
\hline
$\Lambda$CDM
& 1.1596 & 0 & $-1$\\
$w$CDM ($w=-0.8$)
& 1.1536 & $-0.5\%$ & $-0.80$\\
thawing ($\lambda=1.2$)
& 1.1843 & $+2.1\%$ & $-0.77$\\
freezing ($n=1$)
& 1.1277 & $-2.7\%$ & $-0.76$\\
crossing $\nearrow$ ($b=1$)
& 1.2035 & $+3.8\%$ & $-0.80$\\
crossing $\searrow$ ($b=1$)
& 1.1171 & $-3.7\%$ & $-1.20$\\
HDE ($c=0.7$)
& 1.1269 & $-2.8\%$ & $-1.13$\\
GCG ($\alpha=0.5$, $C=0.32$)
& 1.0968 & $-5.4\%$ & $-0.95$\\
DGP
& 0.9685 & $-16.5\%$ & $-0.77$\\
BRANE1 ($K=3$)
& 1.2336 & $+6.4\%$ & $-1.11$\\
BRANE2 ($K=2$)
& 1.0829 & $-6.6\%$ & $-0.90$\\
Cardassian ($n=0.1$)
& 1.2220 & $+5.4\%$ & $-0.90$\\
shift-symm.\ Horndeski ($n=\tfrac14$)
& 1.2335 & $+6.4\%$ & $-1.08$\\
\hline
CPL ($w_0=-0.75$, $w_a=-0.85$)
& 1.1929 & drifts & $-0.75$\\
\end{tabular}
\end{ruledtabular}
\label{tab:models}
\end{table}

\noindent
\textit{\underline{Cosmological models within the universality class}:}

\begin{figure}[t]
\centering
\includegraphics[width=\columnwidth]{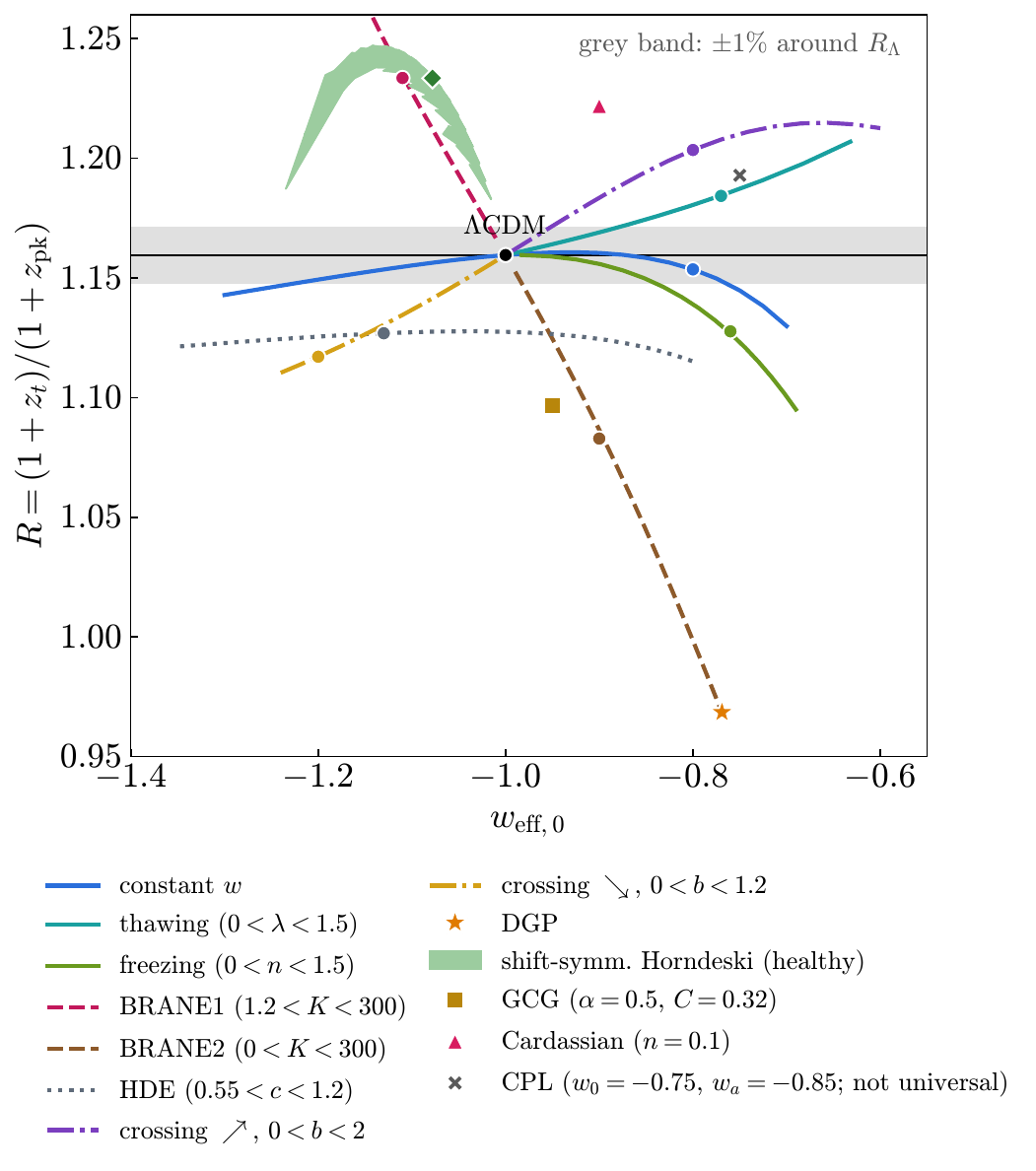}
\caption{$R$ versus $w_{\rm eff,0}$ for the model
families shown, evaluated at $\Omega_{m0}=0.3$.
$R$ itself is independent of $\Omega_{m0}$ for the
universal families;
$w_{\rm eff,0}$ can vary as the present epoch
is changed along a trajectory. For details, see Supplemental Material.}
\label{fig:dist}
\end{figure}

We examine thirteen illustrative cosmological model
families, encompassing constant and evolving dark energy,
phantom-divide crossing, braneworld cosmologies, and
modified gravity. Table~\ref{tab:models} reports their
predictions at $\Omega_{m0}=0.2,0.3,0.4$, showing model-dependent
departures of $R$ from $\RL$, while Fig.~\ref{fig:dist}
shows its dependence on the present effective DE EoS,
$w_{\rm eff,0}$. The Supplemental Material provides
the governing equations, parameter choices, numerical
implementation, and results for different $\Omega_{m0}$,
verifying the invariance of $R$.
The background sets the clock rate via
$|\Om'|=3|w_T|\Om$,
when $\Om$ is monotonic,
while the growth endpoint depends on both the
background and the gravitational clustering strength:
$\mu$ enters the peak condition explicitly and affects
the preceding evolution of $f$.

For dark energy, a constant EoS $w$ simultaneously
shifts both epochs, so $R$ stays within $1\%$ of $\RL$
for $-1.2<w<-0.8$.
For the evolving-EoS families
considered here,
the departure of $R$ from $\RL$ has a definite sign:
$R>\RL$ when $w$ rises between the two epochs
(thawing fields or a crossing from phantom to
quintessence), and $R<\RL$ when it falls
(freezing fields or the opposite crossing).
For the smooth dark energy
models within GR, departures of $R$ from $\RL$ are typically
of order $5\%$ over $-1.3\lesssim w_0\lesssim-0.6$,
a range motivated by recent BAO and SN data%
~\cite{Camilleri:2026rer}.
Models yielding much larger deficits require
substantially less negative $w_0$, potentially placing
them in tension with these data through their
associated expansion histories.

For modified gravity, a change in $G_{\rm eff}$
changes the net pull itself.
At fixed background expansion, this can shift the
growth maximum without changing the onset of
acceleration.
In the examples considered, $R$ shifts by several
percent or more: above $\RL$ for enhanced clustering,
$\mu>1$ (BRANE1 and Cardassian gravity), and below
$\RL$ for weakened clustering, $\mu<1$ (BRANE2 and DGP).
In DGP, the growth peak precedes the onset of acceleration
($R<1$), although its self-accelerating branch is
ghost-unstable and observationally disfavoured.
These examples illustrate the sensitivity of $R$ to
gravity, although the sign of $R-\RL$ does not uniquely
determine whether $\mu>1$ or $\mu<1$ when the background
expansion also changes. Note that in our case $R$ depends solely 
on $\mu$ and is therefore complementary to tests based on weak lensing, 
which probes  $\Sigma$, a combination of two gravitational 
potentials \cite{gallavotti2026a, gallavotti2026b}.

As shown in Table~\ref{tab:models}, $R$ can help break
degeneracies between models with similar present-day
effective EoS.
Thawing quintessence with $\lambda=1.2$, freezing
quintessence with $n=1$, and DGP all have
$w_{\rm eff,0}\simeq-0.77$, yet give
$R=1.1843$, $1.1277$, and $0.9685$, respectively.
Similarly, BRANE1 and HDE can have comparable
$w_{\rm eff,0}$ yet yield $R$
in opposite directions to $\RL$.
The phantom-to-quintessence crossing example,
motivated by recent DESI results~\cite{DESIDR2},
predicts $R\simeq1.20$, providing a concrete target
for future growth measurements.

The diagnostic gains further power when paired
with expansion-only null tests:
flat $\Lambda$CDM has constant
$Om(z)\equiv[H(z)^2/H_0^2-1]/[(1+z)^3-1]$%
~\cite{Sahni2008}, and jerk $j=1$%
~\cite{Visser:2003vq,Sahni2003},
independently of $\Omega_{m0}$.
If the background follows flat $\Lambda$CDM,
with the corresponding conserved clustering-matter
content and standard growing-mode initial conditions,
GR gives $R=\RL$.
Under these assumptions, a significant departure
would directly signal modified growth.
The reference prediction requires no external values
of $\Omega_{m0}$, $H_0$, or $\sigma_{8,0}$.

\smallskip
\noindent
\textit{\underline{$R$-diagnostic:
Current constraints and future prospects}}:
Table~\ref{tab:data} shows the current constraints on
$R$ with available background and growth data.
All combinations give central estimates close to $\RL$: approximately 
$3$--$5\%$ 
above $\RL$ for DESI+SN, and about 
$6$--$7\%$ below ($R\simeq1.07$--$1.09$) 
with the Gold RSD compilation.

At the present precision of 
$\sigma_R/R\simeq9$--$13\%$,
all combinations include $\RL$ within their $68\%$
intervals, providing no significant evidence for a
departure.
Details of the analysis are provided in the
Supplemental Material~\cite{supp}.

Forthcoming data could substantially improve this test.
For the model comparisons considered here,
distinguishing braneworld and DGP gravity from
$\Lambda$CDM at $3\sigma$ requires
$\sigma_{\zpk}\simeq0.015$--$0.08$, while
distinguishing thawing from freezing dark energy
requires $\sigma_{\zpk}\simeq0.002$--$0.013$  
if uncertainty in $z_t$ is negligible.
If the final DESI full-shape data, combined with
low-redshift peculiar velocities, Euclid%
~\cite{Euclid}, and supernovae from the Rubin
Observatory Legacy Survey of Space and Time
(LSST)~\cite{LSST}, bring $\sigma_R/R$ to about
$4$--$6\%$, a deficit of about $10\%$ relative to
$\RL$, 
$\approx 1.5$ times 
to the present central 
deficit
obtained
with Gold-2017 (see Table \ref{tab:data}), 
could be detected at roughly $2$--$3\sigma$.
Such a deficit would exceed the typical
$\simeq5\%$ smooth dark energy shifts 
in the examples considered
and could indicate weaker gravity ($\mu<1$), as in
DGP and BRANE2, subject to consistency with background
expansion constraints\footnote{These are illustrative precision requirements
rather than a survey forecast; a joint forecast must
account for uncertainties in both epochs and their
covariance.}.
Resolving the smaller differences between models
requires percent-level precision, a target for
next-generation spectroscopic surveys.
\smallskip

\begin{table}[t]
\caption{Reconstructed $68\%$ constraints on $z_t$,
$\zpk$, and $R$.
DR2 BAO~\cite{DESIDR2}
+ Ly$\alpha$-AP~\cite{DESIDR2LyaAP} + SNe;
DR1 $\fs$, optionally including Gold-2017~\cite{Gold2017}. PP: Pantheon+~\cite{PantheonPlus};
DES-Dov.: DES-Dovekie~\cite{DESDovekie}.
Planck reference: $\Omega_{m0}=0.3153$
(TT,TE,EE+lowE+lensing)~\cite{Planck2018}.}
\label{tab:data}
\begin{ruledtabular}
\renewcommand{\arraystretch}{1.5}
\setlength{\tabcolsep}{4pt}
\begin{tabular}{lccc}
data & $z_t$ & $\zpk$ & $R$\\
\hline
\multicolumn{4}{l}{DR2 BAO + DR2 Ly$\alpha$ AP + SNe; DR1 $\fs$}\\
\hline
PP
& \multirow{2}{*}{$0.718^{+0.071}_{-0.062}$}
& $0.418^{+0.207}_{-0.176}$
& $1.212^{+0.159}_{-0.182}$\\
PP + Gold
&
& $0.600^{+0.146}_{-0.121}$
& $1.074^{+0.092}_{-0.106}$\\
\hline
DES-Dov.
& \multirow{2}{*}{$0.725^{+0.064}_{-0.060}$}
& $0.415^{+0.198}_{-0.165}$
& $1.219^{+0.149}_{-0.176}$\\
DES-Dov.\ + Gold
&
& $0.593^{+0.148}_{-0.120}$
& $1.082^{+0.091}_{-0.107}$\\
\hline
$\Lambda$CDM ($\Omega_{m0}=0.3$)
& 0.671 & 0.441 & 1.1596\\
$\Lambda$CDM (Planck 2018)
& 0.632 & 0.407 & 1.1596\\
\end{tabular}
\end{ruledtabular}
\end{table}

\noindent
\textit{\underline{Summary}:}
We have established an exact relation between the onset
of cosmic acceleration and the turnover of $\fs$ in flat
$\Lambda$CDM with standard growing-mode evolution:
$\RL=1.1596$, independent of $\Omega_{m0}$ and the overall
expansion and growth normalisations.
We also identify a broad class of dark energy and
modified gravity models in which this independence follows
from a common scaling symmetry and dimensionless trajectory,
at fixed model parameters and early-time initial conditions.
The predicted value of $R$ characterises the expansion
and growth dynamics of each qualifying family. The universality of 
$R$ also holds for interacting dark energy models with coupling of the 
form $Q \propto H \rho_{de}$, 
where matter no longer redshifts as $a^{-3}$; this extension 
will be reported separately.
Current observations are consistent with the $\Lambda$CDM
prediction, while improved growth measurements from ongoing and
next-generation surveys could test departures at the
percent level, making $R$ a sensitive probe of different
expansion and growth dynamics.\\

\begin{acknowledgements}
AAS acknowledges funding from ANRF, Govt.\ of India,
under research grant no.\ CRG/2023/003984.
PM acknowledges funding from the Korea Astronomy and
Space Science Institute (KASI) through the project
``Research on the Principles of the Accelerating
Expansion of the Universe''
(Project Code: 2026183201).

Anthropic Claude (opus 5.5)  has been used for proofreading of the manuscript 
and for numerical checks.The authors have obtained and verified all the scientific contents, derivations and conclusions of the mansucript and take full responsibility 
for all results presented in the article.
\end{acknowledgements}

\bibliographystyle{apsrev4-2}
\bibliography{refs}

\clearpage
\onecolumngrid
\begin{center}
{\large\bfseries Supplemental Material}\\[4pt]
{\large\bfseries A Universal Ratio Linking Cosmic Expansion and the Growth of Structure}\\[6pt]
Anjan A Sen and Purba Mukherjee
\end{center}
\vspace{6pt}
\twocolumngrid
\setcounter{equation}{0}\setcounter{figure}{0}\setcounter{table}{0}\setcounter{section}{0}
\renewcommand{\theequation}{S\arabic{equation}}
\renewcommand{\thefigure}{S\arabic{figure}}
\renewcommand{\thetable}{S\Roman{table}}
\renewcommand{\theHequation}{S\arabic{equation}}
\renewcommand{\theHfigure}{S\arabic{figure}}
\renewcommand{\theHtable}{S\Roman{table}}
\renewcommand{\theHsection}{S\arabic{section}}
\renewcommand{\thesection}{S\Roman{section}}
\setcounter{secnumdepth}{2}
\providecommand{\add}{}\renewcommand{\add}[1]{{\color{blue}#1}}
\providecommand{\del}{}\renewcommand{\del}[1]{\ifmmode\text{\color{red}\sout{$#1$}}\else{\color{red}\sout{#1}}\fi}
\providecommand{\Om}{}\renewcommand{\Om}{\Omega_m}
\providecommand{\fs}{}\renewcommand{\fs}{f\sigma_8}
\providecommand{\zt}{}\renewcommand{\zt}{z_{t}}
\providecommand{\zpk}{}\renewcommand{\zpk}{z_{\rm pk}}
\providecommand{\xpk}{}\renewcommand{\xpk}{x_{\rm pk}}
\providecommand{\Opk}{}\renewcommand{\Opk}{\Omega_{m,\rm pk}}
\providecommand{\fpk}{}\renewcommand{\fpk}{f_{\rm pk}}
\providecommand{\RL}{}\renewcommand{\RL}{R_\Lambda}
\providecommand{\dd}{}\renewcommand{\dd}{\mathrm{d}}

This Supplemental Material presents the background and
linear-growth equations, initial conditions, and numerical
procedure used to obtain the results in Tables~I and II of the
\textit{{\it letter}}. It also describes the determination of the
two characteristic redshifts, $z_t$ and $\zpk$, and the
conditions under which the diagnostic
\begin{equation}
R\equiv\frac{1+z_t}{1+\zpk}=e^{N_{\rm pk}-N_t}
\label{S:eq:Rdef}
\end{equation}
remains independent of $\Omega_{m0}$ when the present epoch
is varied along a common dimensionless trajectory.
Following the notation of the {\it letter}, $N\equiv\ln a$,
and primes denote $\mathrm{d}/\mathrm{d}N$.
We use $\Lambda$CDM to represent a cosmological model with
a cosmological constant $\Lambda$ and pressureless matter,
neglecting radiation in the late-time approximation.
The supplementary tables present results for
$\Omega_{m0}=0.25,0.30,0.35$, complementing those for
$0.20,0.30,0.40$ in Table~I of the {\it letter}.
We use $w_0$ for the present-day dark-energy equation of
state and, in modified-gravity models, its effective
counterpart $w_{\rm eff,0}$.

\section{Evolution equations and numerical methods}\label{S:sec:how}

\subsection{Conditions for universality}

The logarithm of the ratio, $\ln R$, measures the separation
in $e$-folds between the onset of acceleration and the growth 
turnover. It is independent of $\Omega_{m0}$ when changing 
the present matter fraction shifts both epochs equally in 
$\ln a$, leaving their separation unchanged. Physically, the
Universe passes through the same sequence of background and 
growth states, while a different point in this evolution is 
identified as the present epoch.

For the model families considered here, this requires
dimensionless evolution equations with no explicit
dependence on $a$ or $N$ (\textit{no clock}), fixed
dimensionless model parameters (\textit{Lagrangian labels}),
and early-time initial conditions selecting a common
physical history (\textit{a unique history}). The fixed
parameters may include a constant equation of state (EoS),
the slope or power of a scalar potential, or a braneworld 
scale ratio. The initial state may be specified by a frozen 
field, a tracker solution, or an early-time component 
fraction. Autonomous equations alone are insufficient, 
since different initial conditions can select different
background and growth histories. For thawing quintessence, 
the frozen initial condition ($\gamma\to0$, $\Omega_\phi\to0$ 
with fixed $\lambda_i$) uniquely specifies the trajectory; 
for freezing quintessence, the tracker attractor serves the same role.

Along a common trajectory, $\Omega_m$ and the dark energy
EoS have a model-dependent relation. Choosing a different 
point as today changes the present-day matter density 
$\Omega_{m0}$ and may also change the dark energy EoS 
at present-epoch $w_0$, with their relation determined by 
the underlying physical model. Nonetheless, these changes 
preserve the same background and growth history, leaving $R$ 
unchanged. For evolving dark energy, varying $\Omega_{m0}$ 
while holding $w_0$ fixed may instead require different model 
parameters or initial conditions, changing the history
and hence $R$.

\subsection{Linear growth and effective gravitational coupling}\label{S:sec:growth}

We assume separately conserved pressureless
matter (cold dark matter and baryons)
and describe its linear growth on subhorizon scales in the quasi-static approximation.
With the Fourier convention $\nabla^2\to-k^2$,
the effect of the dark sector on gravitational
clustering is encoded in the modified Poisson equation
for $\Psi$:
\begin{equation}
    k^2\Psi=-4\pi G\,a^2\mu\rho_m\delta_m\, .
    \label{S:eq:psi}
\end{equation} Combining this with the linearised continuity and Euler equations gives the growth equation for $D\propto\delta_m$, as
\begin{equation}
D''+(1-q)D'=\frac32\mu\Om D\, .
\label{S:eq:Dgrowth}
\end{equation}
Here $\Om=\frac{8\pi G\rho_m}{3H^2}$ refers to
the pressureless clustering matter and 
$\mu\equiv \frac{G_{\rm eff}}{G}$ specifies the 
effective gravitational coupling.
Once $\mu$ is specified,
gravitational slip does not enter the growth 
equation separately; it affects lensing through
the relation between the two metric potentials.

We set $\mu=1$ for the smooth dark energy (DE)
calculations: constant EoS dark energy ($w$CDM),
quintessence, the phantom-divide crossing models,
holographic dark energy (HDE), and the generalised
Chaplygin gas (GCG). For the crossing, HDE, and GCG models, we adopt the smooth DE approximation, neglecting their perturbations on the scales considered. So, only pressureless matter contributes to the clustering source. This is an additional assumption beyond the specified EoS.

For DGP and the induced-gravity braneworlds with vanishing bulk cosmological constant,
we use the quasistatic brane-bending coupling
$\mu=1+\frac{1}{3\beta}$~\cite{S:Lue2004,S:KoyamaMaartens2006}.
For the Cardassian background, we compare standard 
GR growth ($\mu=1$) with the adopted modified growth
prescription $\mu=\frac{\partial H^2}{\partial(8\pi G\rho_m/3)}$, keeping the background expansion history identical in both cases.
For shift-symmetric Horndeski gravity,
we use the small-scale coupling $\mu=1+\alpha_B^2/(2\alpha c_s^2)$
in the Bellini--Sawicki parametrization,
with $\alpha_M=\alpha_T=0$ for constant $G_4$
~\cite{S:BelliniSawicki2014}.
Here the growth equation is evolved together
with the autonomous background equations.
The corresponding model equations and
coupling functions, including $\beta$, are defined in the following section \ref{S:sec:models} below.

In the calculations presented here, $\mu$ depends on
the background evolution and is taken to be independent of the comoving wavenumber $k$. The quasistatic small-scale approximation requires $k/a\gg H/c_s$, where $c_s$ is the scalar sound speed, together with the other quasistatic
conditions of the theory. These conditions must hold
over the scales and redshifts probed by RSD.
Models with appreciable scale dependence in $\mu$ require a growth factor $D(k,a)$ and are beyond the scope of the present analysis.

\subsection{Initial conditions and characteristic epochs}
\label{S:sec:numerics}

We integrate the growth equation (Eq.~\eqref{S:eq:Dgrowth}) from $a_i=10^{-3}$ for the closed-form and scalar-field backgrounds, and from $\Omega_{{\rm DE},i}=10^{-9}$ for Horndeski gravity, where subscript $i$ denotes the initial epoch.  
For models approaching the Einstein--de Sitter limit
during matter domination, $\Omega_m\to1$ and $\mu\to1$, and the growing mode satisfies $D\propto a$, or equivalently $D'=D$. For the GCG and Cardassian calculations, we initialise
$f=D'/D$ using the positive root of
\begin{equation}
f^2+(1-q)f=\tfrac32\mu\Omega_m,
\end{equation}
with all quantities evaluated at the initial epoch. 
This allows for the early-time clustering matter
fraction in smooth GCG to differ from unity.

The deceleration-acceleration transition redshift $z_t$ is located where $q$ changes from positive to negative as $N$ increases. This corresponds to a minimum of $H(z)/(1+z)$, or equivalently a maximum of $(1+z)D_H(z)$, where $D_H=c/H$. Since $\fs=\sigma_{8,0}D'/D_0$, the
growth-turnover redshift $\zpk$ is located where
$D''$ changes from positive to negative, satisfying
\begin{equation}
(1-q)D'=\tfrac32\mu\Om D \, .
\end{equation}
Thus, $R$ links extrema in the background expansion
and growth histories. Neither $\sigma_{8,0}$ nor the
normalisation of $D$ affects $\zpk$. For the adopted
Cardassian prescription, we select the local maximum
near the acceleration transition. Both epochs are
determined by root finding along the integrated solution. The relative integration tolerance is set to $10^{-11}$. We evaluate each background at the chosen $\Omega_{m0}$,
using analytic $H(a)$ where available and otherwise
integrating the background equations. For quintessence,
the potential amplitude is adjusted to obtain each
$\Omega_{m0}$. We compare the predicted $R$ across different values of $\Omega_{m0}$ at fixed model parameters and initial conditions. For shift-symmetric Horndeski gravity, the present epoch is varied along the same computed trajectory.

\section{Model equations}\label{S:sec:models}

\subsection{$w$CDM}\label{S:sec:wcdm}

For a constant dark-energy EoS $w$, assuming smooth
dark energy,
\begin{equation}
\frac{H^2}{H_0^2}=\Omega_{m0}a^{-3}
+(1-\Omega_{m0})a^{-3(1+w)},
\end{equation}
with $\mu=1$. Defining the density ratio
\begin{equation}
x\equiv\frac{\rho_{\rm DE}}{\rho_m}
=\frac{1-\Omega_{m0}}{\Omega_{m0}}a^{-3w},
\qquad x'=-3wx,
\end{equation}
we obtain
\begin{equation}
\Om=\frac{1}{1+x},\quad
w_T=\frac{wx}{1+x},\quad
q=\frac12(1+3w_T).
\end{equation}
The growth equation becomes
\begin{equation}
-3wx\frac{\dd f}{\dd x}+f^2+(1-q)f
=\frac{3}{2(1+x)}.
\end{equation}
At fixed $w$, these equations and the growing-mode
condition $f\to1$ in the matter era are independent
of $\Omega_{m0}$. Both characteristic epochs therefore
occur at fixed values of $x$. For $w<-1/3$, the acceleration transition satisfies
$x_t=-\frac{1}{1+3w}$, or $\Omega_t=1+\frac{1}{3w}$.
Writing $\xpk(w)$ for the value of $x$ at the growth
maximum gives
\begin{equation}
R(w)=\left(\frac{x_t}{\xpk}\right)^{\frac{1}{3w}}=\left[
\frac{\Omega_t / (1-\Omega_t)}
{\Opk / (1-\Opk)}
\right]^\frac{1}{3\vert w \vert}\,.
\label{S:eq:Rw}
\end{equation}
For $w=-1$, this recovers the flat $\Lambda$CDM result,
$R=(2\xpk)^{1/3}=1.1596$.
Table~\ref{S:tab:w} illustrates the independence of $R$
from $\Omega_{m0}$ at fixed $w$: over
$-1.2<w<-0.8$, $R$ remains within $1\%$ of $\RL$.
\begin{table*}[t]
\centering
\caption{Constant-$w$ backgrounds with smooth dark energy. At fixed $w$, both event redshifts vary with the assigned present matter fraction, while $R$ is unchanged.}
\label{S:tab:w}
\begingroup
\small
\setlength{\tabcolsep}{3pt}
\begin{tabular}{lcccc}
\toprule
Model & $\Omega_{m0}$ & $z_t$ & $\zpk$ & $R$ \\
\midrule
 & $0.25$ & $0.7693$ & $0.5394$ & $1.14933$ \\
$w=-1.2$ & $0.30$ & $0.6500$ & $0.4356$ & $1.14933$ \\
 & $0.35$ & $0.5486$ & $0.3474$ & $1.14933$ \\
\midrule
 & $0.25$ & $0.8284$ & $0.5756$ & $1.16046$ \\
$w=-0.9$ & $0.30$ & $0.6659$ & $0.4355$ & $1.16046$ \\
 & $0.35$ & $0.5308$ & $0.3192$ & $1.16046$ \\
\midrule
 & $0.25$ & $0.8184$ & $0.5763$ & $1.15361$ \\
$w=-0.8$ & $0.30$ & $0.6376$ & $0.4196$ & $1.15361$ \\
 & $0.35$ & $0.4890$ & $0.2908$ & $1.15361$ \\
\bottomrule
\end{tabular}
\endgroup
\end{table*}

\subsection{Thawing quintessence}\label{S:sec:thaw}

For a canonical scalar field, in units $8\pi G=1$,
\begin{equation}
\begin{gathered}
\ddot\phi+3H\dot\phi+V_{,\phi}=0\, ,\\
3H^2=\rho_m+\tfrac12\dot\phi^2+V\, .
\end{gathered}
\end{equation}
Following Scherrer and Sen~\cite{S:ScherrerSen2008}, we define
\begin{equation}
\gamma\equiv1+w_\phi=\frac{\dot\phi^2}{\rho_\phi},\quad
\lambda\equiv-\frac{V_{,\phi}}{V},\quad
\Gamma\equiv\frac{VV_{,\phi\phi}}{V_{,\phi}^2}.
\end{equation}
Choosing the field orientation such that $\phi'\ge0$,
the evolution equations are
\begin{align}
\gamma'&=-3\gamma(2-\gamma)\nonumber+\lambda(2-\gamma)\sqrt{3\gamma\Omega_\phi} \, ,
\nonumber\\
\Omega_\phi'&=3(1-\gamma)\Omega_\phi(1-\Omega_\phi)\, ,
\nonumber\\
\lambda'&=-\sqrt{3\gamma\Omega_\phi}\lambda^2(\Gamma-1)\,.
\label{S:eq:SS}
\end{align}
Here $\Om=1-\Omega_\phi$, $w_T=(\gamma-1)\Omega_\phi$, and $\mu=1$. The exponential potential $V=V_0e^{-\lambda\phi}$ has $\Gamma=1$ and constant slope $\lambda$, while the quadratic potential $V\propto\phi^2$ has $\Gamma=1/2$ and $\lambda=-2/\phi$. For the latter, the positive initial slopes in Table~\ref{S:tab:thaw} correspond to $\phi_i<0$, with the field rolling towards the minimum at $\phi=0$.

The thawing field is initially frozen in the matter era,
with $\gamma\to0$, $\Omega_\phi\to0$, and a fixed initial
slope $\lambda_i$. We integrate the scalar-field equations
to follow its evolution as it begins to roll. At fixed
potential shape and initial state, the amplitude $V_0$
sets the density scale while preserving the dimensionless
trajectory. Changing $\Omega_{m0}$ therefore selects a
different present epoch along the same trajectory:
$w_0$ changes, while $R$ remains fixed.
Table~\ref{S:tab:thaw} illustrates this behaviour for the
exponential and quadratic potentials 
\cite{S:Wetterich1988,S:CaldwellLinder2005}; both give
$R>\RL$ for the parameters shown.
\begin{table*}[t]
\centering
\caption{Thawing quintessence at fixed potential shape and asymptotic initial slope. The potential amplitude is adjusted for each $\Omega_{m0}$.}
\label{S:tab:thaw}
\begingroup
\small
\setlength{\tabcolsep}{3pt}
\begin{tabular}{lccccc}
\toprule
Model & $\Omega_{m0}$ & $z_t$ & $\zpk$ & $w_0$ & $R$ \\
\midrule
 & $0.25$ & $0.8424$ & $0.5820$ & $-0.938$ & $1.16463$ \\
$e^{-\lambda\phi}$, $\lambda=0.6$ & $0.30$ & $0.6855$ & $0.4473$ & $-0.945$ & $1.16463$ \\
 & $0.35$ & $0.5555$ & $0.3356$ & $-0.951$ & $1.16463$ \\
\midrule
 & $0.25$ & $0.9355$ & $0.6343$ & $-0.743$ & $1.18428$ \\
$e^{-\lambda\phi}$, $\lambda=1.2$ & $0.30$ & $0.7328$ & $0.4632$ & $-0.771$ & $1.18428$ \\
 & $0.35$ & $0.5725$ & $0.3278$ & $-0.797$ & $1.18428$ \\
\midrule
 & $0.25$ & $0.8447$ & $0.5835$ & $-0.932$ & $1.16492$ \\
$\phi^2$, $\lambda_i=0.6$ & $0.30$ & $0.6868$ & $0.4480$ & $-0.940$ & $1.16492$ \\
 & $0.35$ & $0.5561$ & $0.3358$ & $-0.947$ & $1.16492$ \\
\midrule
 & $0.25$ & $0.9235$ & $0.6317$ & $-0.761$ & $1.17883$ \\
$\phi^2$, $\lambda_i=1.0$ & $0.30$ & $0.7277$ & $0.4656$ & $-0.799$ & $1.17883$ \\
 & $0.35$ & $0.5736$ & $0.3349$ & $-0.829$ & $1.17883$ \\
\bottomrule
\end{tabular}
\endgroup
\end{table*}

\subsection{Freezing quintessence}\label{S:sec:freeze}

We consider the inverse power-law potential
$V(\phi)=M^{4+n}\phi^{-n}$, where $M$ is a mass scale
and $n>0$ determines the power of the potential.
For this potential, $\Gamma=1+1/n$, and the
matter-era tracker solution satisfies
\begin{equation}
w_\phi=-\frac{2}{n+2}, \quad
\gamma=\frac{n}{n+2}, \quad
\lambda^2\Omega_\phi=3\gamma \quad (\Omega_\phi\ll1).
\label{S:eq:tracker}
\end{equation}
These relations hold in the matter-dominated limit;
the tracker evolves as the field becomes dynamically
important~\cite{S:RatraPeebles1988,S:Steinhardt1999}.
With tracker initial conditions, $n$ fixes the
dimensionless trajectory, while $M$ sets the density
scale. Changing $\Omega_{m0}$ therefore selects a
different present epoch along the same trajectory,
changing $w_0$ but leaving $R$ unchanged.
Table~\ref{S:tab:freeze} illustrates this behaviour,
with $R<\RL$ for the parameters shown.
The initial conditions must place the field on the
tracker before either characteristic epoch.
\begin{table*}[t]
\centering
\caption{Tracker quintessence with $V\propto\phi^{-n}$. The trajectory is selected by its matter-era tracker initial condition.}
\label{S:tab:freeze}
\begingroup
\small
\setlength{\tabcolsep}{3pt}
\begin{tabular}{lccccc}
\toprule
Model & $\Omega_{m0}$ & $z_t$ & $\zpk$ & $w_0$ & $R$ \\
\midrule
 & $0.25$ & $0.8164$ & $0.5680$ & $-0.940$ & $1.15841$ \\
$n=0.2$ & $0.30$ & $0.6613$ & $0.4342$ & $-0.937$ & $1.15841$ \\
 & $0.35$ & $0.5315$ & $0.3221$ & $-0.934$ & $1.15841$ \\
\midrule
 & $0.25$ & $0.8004$ & $0.5640$ & $-0.867$ & $1.15112$ \\
$n=0.5$ & $0.30$ & $0.6340$ & $0.4194$ & $-0.860$ & $1.15112$ \\
 & $0.35$ & $0.4950$ & $0.2988$ & $-0.853$ & $1.15112$ \\
\midrule
 & $0.25$ & $0.7454$ & $0.5477$ & $-0.775$ & $1.12774$ \\
$n=1.0$ & $0.30$ & $0.5652$ & $0.3879$ & $-0.763$ & $1.12774$ \\
 & $0.35$ & $0.4156$ & $0.2552$ & $-0.752$ & $1.12774$ \\
\bottomrule
\end{tabular}
\endgroup
\end{table*}

\subsection{DGP gravity}\label{S:sec:dgp}
We consider the spatially flat self-accelerating branch
of Dvali--Gabadadze--Porrati (DGP) gravity~\cite{S:DGP2000},
\begin{equation}
H^2-\frac{H}{r_c}=\frac{8\pi G}{3}\rho_m\, ,
\label{S:eq:dgp}
\end{equation}
where $r_c$ is the crossover length. The subhorizon
quasistatic coupling is~\cite{S:Lue2004,S:KoyamaMaartens2006}
\begin{equation}
\mu=1+\frac{1}{3\beta}\, ,  \qquad 
\beta=1-2r_cH\left(1+\frac{\dot H}{3H^2}\right).
\end{equation}
Defining $h\equiv1/(r_cH)$ gives $1-h=\Om$.
Matter conservation then yields
$H'/H=-3\Om(1+\Om)^{-1}$, from which
\begin{equation}
\begin{gathered}
w_T=-\frac{1-\Om}{1+\Om},\\
\beta=-\frac{1+\Om^2}{1-\Om^2},\\
\mu=1-\frac{1-\Om^2}{3(1+\Om^2)}.
\label{S:eq:dgpclosure}
\end{gathered}
\end{equation}
Together with $\Om'=3w_T\Om$ and the matter-era
growing-mode condition, these equations determine
a common dimensionless trajectory. Flatness fixes
$r_cH_0=(1-\Omega_{m0})^{-1}$, while $r_c$ sets
the overall density scale, proportional to
$1/(Gr_c^2)$. Neither $r_c$ nor $\Omega_{m0}$ enters
the dimensionless evolution equations. Consequently, every member of this family has
\begin{equation}
R_{\rm DGP}=0.9685.
\end{equation}
Since $R<1$, the growth turnover precedes the onset
of acceleration. Changing $\Omega_{m0}$ shifts both
epochs and changes
$w_{\rm eff,0}=-1/(1+\Omega_{m0})$, while leaving
$R$ unchanged (Table~\ref{S:tab:dgp}).
\begin{table*}[t]
\centering
\caption{The self-accelerating DGP background with the quasistatic growth coupling. Here $w_0=w_{\rm eff,0}$. The maximum precedes the acceleration transition.}
\label{S:tab:dgp}
\begingroup
\small
\setlength{\tabcolsep}{3pt}
\begin{tabular}{lcccccc}
\toprule
Model & $\Omega_{m0}$ & $z_t$ & $\zpk$ & $w_0$ & $\mu_{\rm pk}$ & $R$ \\
\midrule
 & $0.25$ & $0.6510$ & $0.7046$ & $-0.800$ & $0.8068$ & $0.96854$ \\
DGP & $0.30$ & $0.4838$ & $0.5320$ & $-0.769$ & $0.8068$ & $0.96854$ \\
 & $0.35$ & $0.3415$ & $0.3851$ & $-0.741$ & $0.8068$ & $0.96854$ \\
\bottomrule
\end{tabular}
\endgroup
\end{table*}

\subsection{Induced-gravity braneworlds}\label{S:sec:brane}

We consider the Sahni--Shtanov braneworld with vanishing
bulk cosmological constant and dark radiation
\cite{S:SahniShtanov2003}. With brane tension $\sigma$, the background equations are,
\begin{align}
H^2-\frac{\varepsilon H}{r_c}
&=\frac{8\pi G}{3}(\rho_m+\sigma)\, ,\nonumber\\
\frac{H}{H_0}
&=\varepsilon\sqrt{\Omega_l}
+\sqrt{\Omega_l+\Omega_{m0}a^{-3}+\Omega_\sigma}\,.
\label{S:eq:brane}
\end{align}
The branches $\varepsilon=-1$ and $\varepsilon=+1$
correspond to BRANE1 and BRANE2, respectively, with
\begin{equation}
\begin{gathered}
\Omega_l=\frac{1}{4r_c^2H_0^2}\,,\qquad
\Omega_\sigma=\frac{8\pi G\sigma}{3H_0^2}\,,\\
1=\Omega_{m0}+\Omega_\sigma
+2\varepsilon\sqrt{\Omega_l}\,.
\end{gathered}
\end{equation}
For growth in this induced-gravity limit, we adopt
\begin{equation}
\mu=1+\frac{1}{3\beta}\, , \qquad 
\beta=1-2\varepsilon r_cH
\left(1+\frac{\dot H}{3H^2}\right).
\end{equation}
Defining $h\equiv(r_cH)^{-1}$ and the dimensionless scale ratio
\begin{equation}
K\equiv\frac{8\pi G}{3}\sigma r_c^2
=\frac{\Omega_\sigma}{4\Omega_l},
\label{S:eq:K}
\end{equation}
we obtain
\begin{equation}
\begin{aligned}
\Om&=1-\varepsilon h-Kh^2\, , \quad
w_T=-1+\frac{2\Om}{2-\varepsilon h}\, ,\\
h'&=\frac{3h\Om}{2-\varepsilon h}\, , \qquad 
\beta=1-\frac{2\varepsilon(1+Kh^2)}
{h(2-\varepsilon h)}\, .
\end{aligned}
\label{S:eq:braneclosure}
\end{equation}
At fixed branch and $K$, these equations and the
matter-era growing-mode condition determine a common
dimensionless trajectory. Changing $\Omega_{m0}$
therefore shifts the two epochs equally, leaving
$R$ unchanged. At fixed $\Omega_l$, flatness gives
\begin{equation}
K=\frac{1-\Omega_{m0} -2\varepsilon\sqrt{\Omega_l}}{4\Omega_l}\, .
\end{equation}
Thus, changing $\Omega_{m0}$ changes $K$ and the
trajectory, so $R$ can vary. Table~\ref{S:tab:brane} compares both choices.
Both branches approach $\Lambda$CDM as $K\to\infty$,
while BRANE2 approaches DGP as $K\to0$.
Since $\Om$ is not always monotonic on BRANE1,
we describe the evolution using $h=1/(r_cH)$. 
The non-monotonic phase ($\Om>1$ for $h<1/K$) occurs only at early times: $w_T<0$ requires $h>1/(2K)$, which holds from the acceleration transition onward since $h$ increases with time. Hence $\Om$ decreases monotonically between the two epochs, as assumed in Eq.~(7) of the {\it letter}.
\begin{table*}[t]
\centering
\caption{Induced-gravity braneworlds. At fixed branch and $K=(8\pi G/3)\sigma r_c^2$, the dimensionless trajectory is unchanged. Fixed $\Omega_l$ instead changes $K$ when $\Omega_{m0}$ varies. Here $w_0=w_{\rm eff,0}$.}
\label{S:tab:brane}
\begingroup
\small
\setlength{\tabcolsep}{3pt}
\begin{tabular}{lcccccc}
\toprule
Model & $\Omega_{m0}$ & $z_t$ & $\zpk$ & $w_0$ & $\mu_{\rm pk}$ & $R$ \\
\midrule
 & $0.25$ & $0.9476$ & $0.5788$ & $-1.086$ & $1.0898$ & $1.23360$ \\
BRANE1, $K=3$ & $0.30$ & $0.8044$ & $0.4627$ & $-1.108$ & $1.0898$ & $1.23360$ \\
 & $0.35$ & $0.6859$ & $0.3667$ & $-1.134$ & $1.0898$ & $1.23360$ \\
\midrule
 & $0.25$ & $0.8811$ & $0.5686$ & $-1.047$ & $1.0496$ & $1.19926$ \\
BRANE1, $K=10$ & $0.30$ & $0.7373$ & $0.4486$ & $-1.059$ & $1.0496$ & $1.19926$ \\
 & $0.35$ & $0.6175$ & $0.3488$ & $-1.072$ & $1.0496$ & $1.19926$ \\
\midrule
 & $0.25$ & $0.9033$ & $0.5713$ & $-1.061$ & $1.0641$ & $1.21125$ \\
BRANE1, $\Omega_l=0.05$ & $0.30$ & $0.7620$ & $0.4533$ & $-1.078$ & $1.0654$ & $1.21241$ \\
 & $0.35$ & $0.6453$ & $0.3557$ & $-1.098$ & $1.0669$ & $1.21364$ \\
\midrule
 & $0.25$ & $0.7299$ & $0.5975$ & $-0.914$ & $0.9089$ & $1.08292$ \\
BRANE2, $K=2$ & $0.30$ & $0.5769$ & $0.4562$ & $-0.895$ & $0.9089$ & $1.08292$ \\
 & $0.35$ & $0.4472$ & $0.3364$ & $-0.877$ & $0.9089$ & $1.08292$ \\
\midrule
 & $0.25$ & $0.7698$ & $0.5765$ & $-0.957$ & $0.9543$ & $1.12259$ \\
BRANE2, $K=10$ & $0.30$ & $0.6207$ & $0.4438$ & $-0.947$ & $0.9543$ & $1.12259$ \\
 & $0.35$ & $0.4950$ & $0.3318$ & $-0.937$ & $0.9543$ & $1.12259$ \\
\midrule
 & $0.25$ & $0.7443$ & $0.5881$ & $-0.930$ & $0.9260$ & $1.09834$ \\
BRANE2, $\Omega_l=0.03$ & $0.30$ & $0.5890$ & $0.4516$ & $-0.910$ & $0.9218$ & $1.09466$ \\
 & $0.35$ & $0.4554$ & $0.3349$ & $-0.887$ & $0.9169$ & $1.09020$ \\
\bottomrule
\end{tabular}
\endgroup
\end{table*}

\subsection{Phantom-divide crossing models}\label{S:sec:cross}

For an independently conserved DE component with
$w=w(\Omega_{\rm DE})$, the background equations are
\begin{equation}
\begin{aligned}
\Omega_{\rm DE}'&=-3w(\Omega_{\rm DE})
\Omega_{\rm DE}(1-\Omega_{\rm DE})\, ,\\
w_T&=w(\Omega_{\rm DE})\Omega_{\rm DE}\, .
\end{aligned}
\label{S:eq:stateDE}
\end{equation}
At fixed model parameters, these equations with $\mu=1$
and matter-era growing-mode initial conditions describe
the evolution equations for different
values of $\Omega_{m0}$ along a given monotonic branch.
A crossing occurs at a fixed DE fraction
$\Omega_{\rm DE}=\Omega_\times$, where
$w(\Omega_\times)=-1$.

\noindent We consider the phenomenological prescription \begin{equation}
w=-1\pm b\left(\Omega_{\rm DE}-\frac12\right)\, ,
\label{S:eq:wlin}
\end{equation}
which crosses $w=-1$ at $\Omega_\times=1/2$.
For the positive $b$ values,
$\Omega_{\rm DE}$ increases: the upper sign gives
a phantom-to-quintessence crossing, while
the lower sign gives the reverse. At fixed $b$,
$R$ remains independent of $\Omega_{m0}$. 
The rising and falling examples give $R>\RL$ and
$R<\RL$, respectively. For the rising branch with
$b=1$, we obtain $R=1.2035$,
$w(z_t)\simeq-1.23$, and $w(\zpk)\simeq-1.09$.

Interestingly, applying the same growth prescription
to the fitted logistic background of
Ref.~\cite{S:GuptaChoudhury:2026hwk} gives
$R\simeq1.20$, for comparison. The CPL parametrization $w=w_0+w_a(1-a)$
~\cite{S:CPL2001,S:Linder2003} instead fixes the crossing at $a_\times=1+(1+w_0)/w_a$, provided $w_a\ne0$ and the crossing lies within the evolution considered. Holding $(w_0,w_a)$ fixed while varying $\Omega_{m0}$ generally changes the dimensionless trajectory, so $R$ need not remain fixed.

\subsection{Holographic dark energy}\label{S:sec:hde}

Holographic dark energy (HDE) provides another example
of an EoS determined by $\Omega_{\rm DE}$.
With the future event horizon as the infrared
cutoff~\cite{S:Li2004},
\begin{equation}
w=-\frac13-\frac{2}{3c}\sqrt{\Omega_{\rm DE}}\, ,
\end{equation}
where $c$ is a dimensionless model parameter.
The background follows Eq.~\eqref{S:eq:stateDE}.
For $0<c<1$, the EoS crosses from $w>-1$ to $w<-1$
at $\Omega_\times=c^2$. With $c=0.7$ and DE
perturbations neglected, we obtain $R=1.1269$,
independently of $\Omega_{m0}$.
Table~\ref{S:tab:cross} presents the crossing and HDE examples considered in this work.
\begin{table*}[t]
\centering
\caption{State-dependent dark-energy histories with the smooth-growth prescription. The rising and falling branches refer to Eq.~\eqref{S:eq:wlin}. The crossing occurs at a fixed component fraction, while its redshift changes with $\Omega_{m0}$.}
\label{S:tab:cross}
\begingroup
\small
\setlength{\tabcolsep}{3pt}
\begin{tabular}{lcccccc}
\toprule
Model & $\Omega_{m0}$ & $z_\times$ & $z_t$ & $\zpk$ & $w_0$ & $R$ \\
\midrule
 & $0.25$ & $0.419$ & $0.6948$ & $0.5039$ & $-1.158$ & $1.12692$ \\
HDE, $c=0.7$ & $0.30$ & $0.319$ & $0.5751$ & $0.3977$ & $-1.130$ & $1.12692$ \\
 & $0.35$ & $0.232$ & $0.4713$ & $0.3056$ & $-1.101$ & $1.12692$ \\
\midrule
 & $0.25$ & $0.384$ & $0.6495$ & $0.4766$ & $-1.250$ & $1.11713$ \\
Falling, $b=1$ & $0.30$ & $0.293$ & $0.5405$ & $0.3790$ & $-1.200$ & $1.11713$ \\
 & $0.35$ & $0.212$ & $0.4439$ & $0.2925$ & $-1.150$ & $1.11713$ \\
\midrule
 & $0.25$ & $0.529$ & $1.0540$ & $0.7067$ & $-0.750$ & $1.20350$ \\
Rising, $b=1$ & $0.30$ & $0.372$ & $0.8434$ & $0.5317$ & $-0.800$ & $1.20350$ \\
 & $0.35$ & $0.251$ & $0.6809$ & $0.3966$ & $-0.850$ & $1.20350$ \\
\midrule
 & $0.25$ & $0.676$ & $1.3583$ & $0.9448$ & $-0.500$ & $1.21262$ \\
Rising, $b=2$ & $0.30$ & $0.438$ & $1.0237$ & $0.6689$ & $-0.600$ & $1.21262$ \\
 & $0.35$ & $0.279$ & $0.7995$ & $0.4840$ & $-0.700$ & $1.21262$ \\
\bottomrule
\end{tabular}
\endgroup
\end{table*}

\subsection{Generalised Chaplygin gas}\label{S:sec:gcg}

We consider the generalised Chaplygin gas (GCG) as
a DE component, with EoS and density evolution~\cite{S:Bento2002},
\begin{equation}
\begin{gathered}
p_{\rm GCG}=-{A}/{\rho_{\rm GCG}^{\alpha}}\, ,~  
\rho_{\rm GCG}
=\left[A+B a^{-3(1+\alpha)}\right]^\frac{1}{(1+\alpha)}.
\end{gathered}
\end{equation} where $A,B>0$ and $\alpha>-1$.
The density scale is $\rho_*=A^{1/(1+\alpha)}$.
At early times, GCG behaves like dust, with a fixed
density ratio to matter,
\begin{equation}
C\equiv\lim_{a\to0}\frac{\rho_{\rm GCG}}{\rho_m}
=\frac{B^{1/(1+\alpha)}}{\rho_{m0}}.
\label{S:eq:C}
\end{equation}
We neglect GCG perturbations and set $\mu=1$, so only
the pressureless matter clusters.
The early-time 
background therefore has $q\to1/2$ and
$\Om\to(1+C)^{-1}$, giving the growing solution
$D\propto a^{p_+}$ with
\begin{equation}
p_+=\frac{-1 + \sqrt{1+24(1+C)^{-1}}}{4}.
\label{S:eq:gcginit}
\end{equation}
For fixed values of $(\alpha,C)$, the background and growing-mode
initial conditions select a common dimensionless
trajectory, leaving $R$ independent of $\Omega_{m0}$. 
The conventional present-day parameter $A_0\equiv A/(A+B)$ satisfies
\begin{equation}
A_0=1-\left(
\frac{C \Omega_{m0}}{1-\Omega_{m0}}
\right)^{1+\alpha}.
\label{S:eq:As}
\end{equation}
At fixed $A_0$, changing $\Omega_{m0}$ changes $C$
and the trajectory, so $R$ can vary. Table~\ref{S:tab:gcg} compares both choices.
For $\alpha=0.5$ and $C=0.32$, we obtain
$R=1.0968$, a $5.4\%$ deficit relative to $\RL$.
\begin{table*}[t]
\centering
\caption{GCG treated as a smooth component in addition to clustering pressureless matter. The growing initial index is given by Eq.~\eqref{S:eq:gcginit}. Fixing $C$ preserves the trajectory; fixing $A_0$ does not.}
\label{S:tab:gcg}
\begingroup
\small
\setlength{\tabcolsep}{3pt}
\begin{tabular}{lcccccc}
\toprule
Model & $\Omega_{m0}$ & $z_t$ & $\zpk$ & $w_0$ & $\mu_{\rm pk}$ & $R$ \\
\midrule
 & $0.25$ & $0.6419$ & $0.4970$ & $-0.965$ & $1.0000$ & $1.09681$ \\
$\alpha=0.5$, $C=0.32$ & $0.30$ & $0.5044$ & $0.3716$ & $-0.949$ & $1.0000$ & $1.09681$ \\
 & $0.35$ & $0.3873$ & $0.2649$ & $-0.928$ & $1.0000$ & $1.09681$ \\
\midrule
 & $0.25$ & $0.5935$ & $0.4740$ & $-0.950$ & $1.0000$ & $1.08108$ \\
$\alpha=0.5$, $A_0=0.95$ & $0.30$ & $0.5062$ & $0.3725$ & $-0.950$ & $1.0000$ & $1.09744$ \\
 & $0.35$ & $0.4250$ & $0.2834$ & $-0.950$ & $1.0000$ & $1.11030$ \\
\bottomrule
\end{tabular}
\endgroup
\end{table*}

\subsection{Cardassian expansion}\label{S:sec:card}

We consider the power-law Cardassian model
\cite{S:FreeseLewis2002},
\begin{equation}
H^2=\frac{8\pi G}{3}\rho_m
\left[1+\left(\frac{\rho_m}{\rho_{\rm car}}\right)^{n-1}\right],
\end{equation}
where $\rho_{\rm car}$ sets the density scale and
$n<2/3$ permits late-time acceleration. Its expansion
history,
\begin{equation}
\frac{H^2}{H_0^2}
=\Omega_{m0}a^{-3}+(1-\Omega_{m0})a^{-3n},
\end{equation}
is identical to that of $w$CDM with $w=n-1$.

At the same background expansion, we compare standard
GR growth ($\mu=1$) with the adopted prescription
\begin{equation}
\mu=\frac{\partial H^2}{\partial(8\pi G\rho_m/3)}
=1+n\frac{1-\Om}{\Om},
\label{S:eq:cardmu}
\end{equation}
where model parameters are held fixed in the derivative. This is an additional growth assumption: the background equation alone does not determine $\mu$. At fixed $n$, either prescription gives $R$ independent of $\Omega_{m0}$.

For $n=0.1$, the derivative prescription gives
$R=$ $1.2220$, compared with $1.1605$ for $\mu=1$
(Table~\ref{S:tab:card}). This $5.3\%$ difference comes
entirely from the growth prescription. For the derivative prescription, the quoted turnover
is a local maximum near the acceleration transition.
The far-future solution $D\propto a^p$ satisfies
\begin{equation}
p^2+\left(2-\frac32n\right)p-\frac32n=0.
\end{equation}
For $0<n<2/3$, the growing root has $p>0$,
so $D'$ eventually increases again.
\begin{table*}[t]
\centering
\caption{Cardassian background with two growth prescriptions. The derivative prescription is Eq.~\eqref{S:eq:cardmu}; the second case is smooth $w$CDM with $w=-0.9$. The quoted peak for the derivative prescription is a local maximum near the acceleration transition.}
\label{S:tab:card}
\begingroup
\small
\setlength{\tabcolsep}{3pt}
\begin{tabular}{lcccccc}
\toprule
Model & $\Omega_{m0}$ & $z_t$ & $\zpk$ & $w_0$ & $\mu_{\rm pk}$ & $R$ \\
\midrule
 & $0.25$ & $0.8284$ & $0.4962$ & $-0.900$ & $1.1011$ & $1.22198$ \\
$n=0.1$, derivative $\mu$ & $0.30$ & $0.6659$ & $0.3633$ & $-0.900$ & $1.1011$ & $1.22198$ \\
 & $0.35$ & $0.5308$ & $0.2527$ & $-0.900$ & $1.1011$ & $1.22198$ \\
\midrule
 & $0.25$ & $0.8284$ & $0.5756$ & $-0.900$ & $1.0000$ & $1.16046$ \\
$n=0.1$, $\mu=1$ & $0.30$ & $0.6659$ & $0.4355$ & $-0.900$ & $1.0000$ & $1.16046$ \\
 & $0.35$ & $0.5308$ & $0.3192$ & $-0.900$ & $1.0000$ & $1.16046$ \\
\bottomrule
\end{tabular}
\endgroup
\end{table*}

\begin{figure*}[t]
\centering
\includegraphics[width=0.85\textwidth]{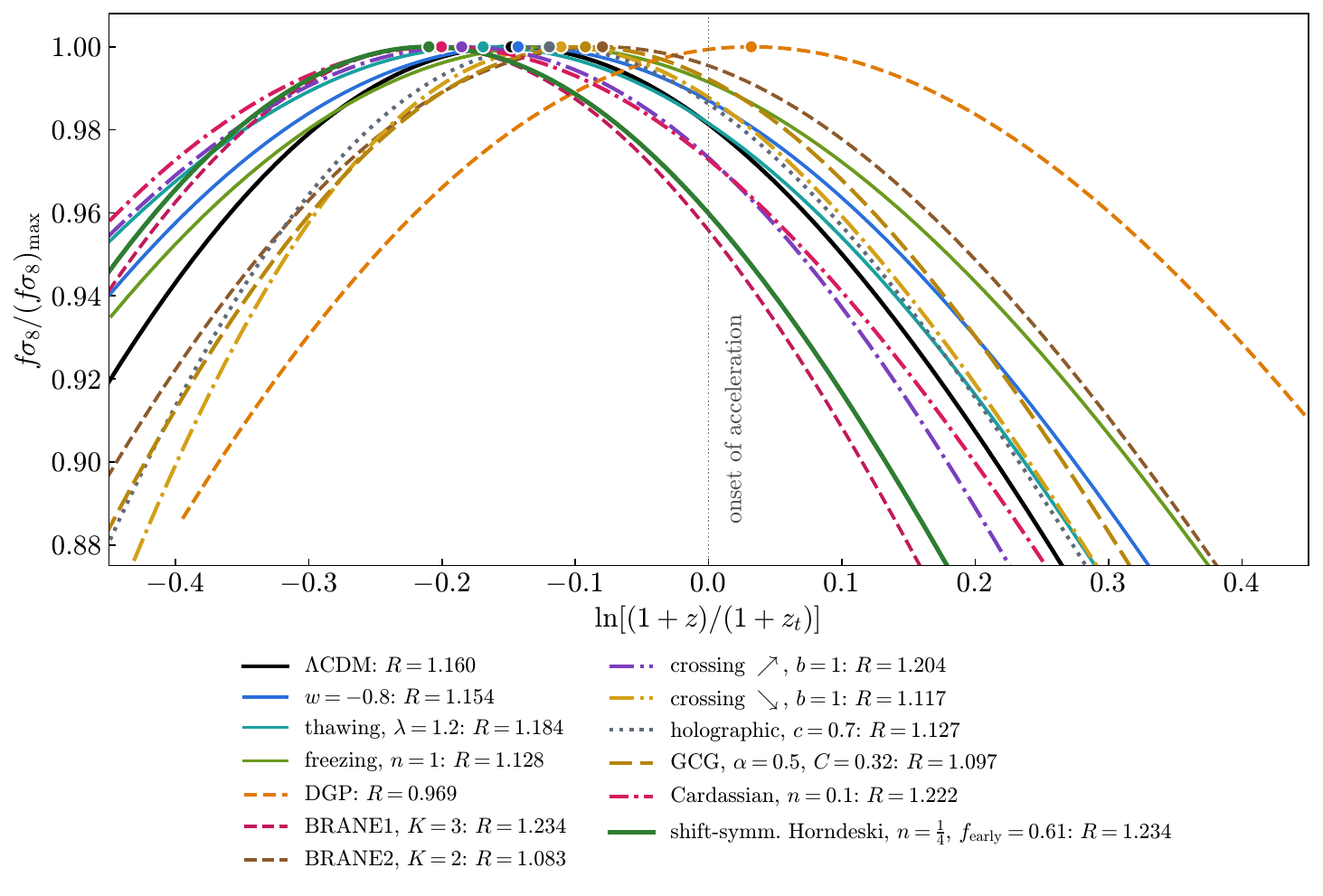}
\caption{Normalised $\fs$ curves aligned at the
acceleration transition, $u=0$ (vertical dotted line).
Circles mark the growth maxima at $u_{\rm pk}=-\ln R$;
cosmic time increases towards the left.
Model parameters and $R$ values are given in the legend.}
\label{S:fig:curves}
\end{figure*}

\subsection{Shift-symmetric Horndeski gravity}\label{S:sec:horn}

We consider the shift-symmetric Horndeski model
studied in Ref.~\cite{S:CalderonLinder2026}, with
kinetic gravity braiding~\cite{S:Horndeski1974,S:Deffayet2010}.
In units $8\pi G=1$, the Lagrangian is
\begin{equation}
\mathcal L=\frac12\mathcal R+K(X)-G_3(X)\Box\phi,
\end{equation}
where $\mathcal R$ is the Ricci scalar and
\begin{equation}
X=\tfrac12\dot\phi^2,\qquad
K\propto X^n,\qquad
G_3\propto X^{(2n-1)/4}.
\end{equation}
The model contains neither a scalar potential nor
a cosmological constant. With $m\equiv(2n-1)/4$, we introduce the dimensionless
variables
\begin{equation}
\begin{gathered}
x_k\equiv\frac{K}{3H^2}\,,\quad
x_g\equiv\frac{6H\dot\phi XG_{3X}}{3H^2}\, ,\quad
x_\phi\equiv\frac{\ddot\phi}{3H\dot\phi}\, , 
\end{gathered}
\end{equation}
where $G_{3X}\equiv\frac{\dd G_3}{\dd X}$. The background equations satisfy
\begin{equation}
\begin{aligned}
\Omega_{\rm DE}&=(2n-1)x_k+x_g\, ,\\
w_T&=x_k-x_gx_\phi\, ,\\
\alpha x_\phi&=-6nx_k-\tfrac32x_g(1-x_k)\, ,
\end{aligned}
\end{equation}
with
\begin{equation}
\begin{gathered}
\alpha=\alpha_K+\tfrac32\alpha_B^2\, ,\\
\alpha_K=6n(2n-1)x_k+6mx_g\, ,\qquad
\alpha_B=x_g\, .
\end{gathered}
\end{equation}
The autonomous evolution equations are
\begin{align}
x_k'&=x_k(6nx_\phi-x_H)\, ,\nonumber\\
x_g'&=x_g\left[\tfrac32(2n+1)x_\phi-\tfrac12x_H\right]\, ,
\label{S:eq:hornsys}
\end{align}
where $x_H\equiv-3(1+w_T)=2H'/H$.
We impose the scalar no-ghost and gradient-stability
conditions
\begin{equation}
\begin{gathered}
\alpha>0\, ,\\
\alpha c_s^2=\tfrac12(2-x_g)(x_g-x_H)+x_g'
-3(1-\Omega_{\rm DE})>0\, .
\label{S:eq:stability}
\end{gathered}
\end{equation}
With constant $G_4=1/2$, $\alpha_M=\alpha_T=0$,
and the quasistatic coupling is~\cite{S:BelliniSawicki2014} \begin{equation}
\mu=1+\frac{\alpha_B^2}{2\alpha c_s^2}\, .
\end{equation}
Each trajectory is labelled by $n$ and the early-time braiding fraction
$f_{\rm early}\equiv\lim_{\Omega_{\rm DE}\to0}
x_g/\Omega_{\rm DE}$. At
$\Omega_{{\rm DE},i}={\eta}\ll1$, we initialise
$x_{g,i}=f_{\rm early} {\eta},$ 
$x_{k,i}=\frac{(1-f_{\rm early}) {\eta}}{2n-1},$
with $f\to1$ initially and
$w_{\rm DE}\to1/(2n-1)<-1$ for $0<n<1/2$. At fixed $(n,f_{\rm early})$, changing $\Omega_{m0}$
selects a different present epoch along the same
trajectory, leaving $R$ unchanged. Table~\ref{S:tab:horn} presents the resulting values of $R$. The tabulated models give $1.2\lesssim R\lesssim1.2471$, all above $\RL$, with departures determined by both the background evolution and the growth history. To map the healthy family as a whole, we scan $0.075\le n\le0.475$ and, for each $n$,
we find the range of $f_{\rm early}$ for which the no ghost and gradient stability conditions
[Eq.~\eqref{S:eq:stability}] hold throughout the evolution. Since each healthy model is specified by the
two constants $(n,f_{\rm early})$, the family occupies a region, not a curve, in
the $(w_{\rm eff,0},R)$ plane; this is the shaded region in Fig.~2 in the main paper. Across
the 156 healthy models, $w_{\rm eff,0}$ ranges from $-1.26$ to $-1.02$ and $R$ from $1.147$
to $1.247$, with a median of $1.227$; $83\%$ of the models have $R>1.20$. The full healthy
family therefore has $1.15\lesssim R\lesssim1.25$, and the trajectories mentioned in
Table~\ref{S:tab:horn} lie within this region.
\begin{table*}[t]
\centering
\caption{Selected shift-symmetric Horndeski trajectories. The tabulated solutions satisfy the stated scalar stability tests over the integration interval. Here $w_0=w_{\rm eff,0}$. Different present epochs are assigned along the same autonomous trajectory.}
\label{S:tab:horn}
\begingroup
\small
\setlength{\tabcolsep}{3pt}
\begin{tabular}{lcccccc}
\toprule
Model & $\Omega_{m0}$ & $z_t$ & $\zpk$ & $w_0$ & $\mu_{\rm pk}$ & $R$ \\
\midrule
 & $0.25$ & $0.8421$ & $0.5351$ & $-1.026$ & $1.0662$ & $1.20001$ \\
$n=1/8$, $f_{\rm early}=0.35$ & $0.30$ & $0.6980$ & $0.4150$ & $-1.032$ & $1.0662$ & $1.20001$ \\
 & $0.35$ & $0.5777$ & $0.3147$ & $-1.038$ & $1.0662$ & $1.20001$ \\
\midrule
 & $0.25$ & $0.8751$ & $0.5169$ & $-1.086$ & $1.1464$ & $1.23615$ \\
$n=1/4$, $f_{\rm early}=0.56$ & $0.30$ & $0.7373$ & $0.4054$ & $-1.108$ & $1.1464$ & $1.23615$ \\
 & $0.35$ & $0.6231$ & $0.3130$ & $-1.131$ & $1.1464$ & $1.23615$ \\
\midrule
 & $0.25$ & $0.8722$ & $0.5178$ & $-1.063$ & $1.1192$ & $1.23350$ \\
$n=1/4$, $f_{\rm early}=0.61$ & $0.30$ & $0.7312$ & $0.4035$ & $-1.078$ & $1.1192$ & $1.23350$ \\
 & $0.35$ & $0.6141$ & $0.3086$ & $-1.095$ & $1.1192$ & $1.23350$ \\
\midrule
 & $0.25$ & $0.8642$ & $0.5162$ & $-1.055$ & $1.1135$ & $1.22956$ \\
$n=1/4$, $f_{\rm early}=0.65$ & $0.30$ & $0.7228$ & $0.4012$ & $-1.069$ & $1.1135$ & $1.22956$ \\
 & $0.35$ & $0.6052$ & $0.3055$ & $-1.084$ & $1.1135$ & $1.22956$ \\
\midrule
 & $0.25$ & $0.9037$ & $0.5265$ & $-1.113$ & $1.1396$ & $1.24708$ \\
$n=3/8$, $f_{\rm early}=0.82$ & $0.30$ & $0.7673$ & $0.4171$ & $-1.142$ & $1.1396$ & $1.24708$ \\
 & $0.35$ & $0.6548$ & $0.3270$ & $-1.174$ & $1.1396$ & $1.24708$ \\
\bottomrule
\end{tabular}
\endgroup
\end{table*}

\section{Normalised growth curves}\label{S:sec:curves}

To compare the growth histories, we align the onset
of late-time acceleration and normalise each curve
to its peak value $\fs(\zpk)$. We define
\begin{equation}
\begin{gathered}
u\equiv\ln\frac{1+z}{1+z_t}=N_t-N,\\
\widehat{\fs}(u)\equiv\frac{\fs(u)}{\fs(u_{\rm pk})}.
\end{gathered}
\end{equation}
The late-time transition from decelerated to accelerated expansion occurs at $u=0$. Since
$R=(1+z_t)/(1+\zpk)$, the growth rate turnover peak lies at $u_{\rm pk}=-\ln R$. 
Negative $u$ therefore corresponds to epochs after
the transition $z_t$.  
For each universal family, the normalised curve is independent of $\Omega_{m0}$, at fixed model parameters and initial conditions: shifting the present epoch along a common trajectory leaves the normalised curve unchanged.

Figure~\ref{S:fig:curves} compares the representative
models summarized in Sec. \ref{S:sec:models}. A larger $R$ places the peak further to the left, indicating a longer separation in $e$-folds after the onset of acceleration. DGP instead has $R<1$ and peaks at $u_{\rm pk}>0$: its weakened gravitational clustering, together with its background evolution, causes the growth turnover to precede acceleration. The other displayed models peak after the transition.

\begin{figure*}    
\includegraphics[width=0.75\textwidth]{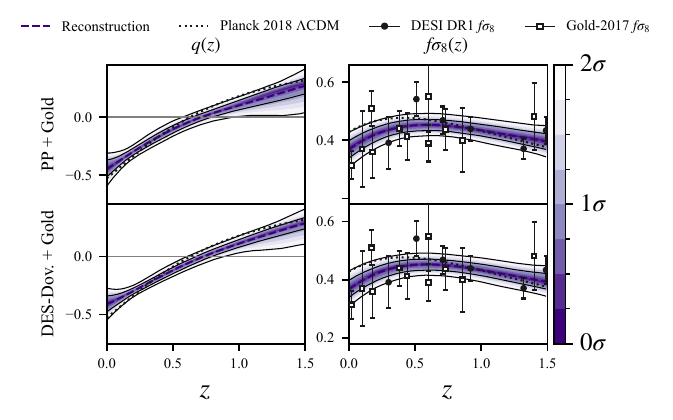}
\caption{Reconstructed deceleration parameter $q(z)$ (left) and
    growth rate $\fs(z)$ (right) for DR2 BAO + SNIa with DESI DR1
    $\fs$ data (top: PP + Gold; bottom: DES-Dov.\ + Gold).} 
    \label{S:fig:recz}
\end{figure*}

\begin{figure*}    
\includegraphics[width=0.6\textwidth]{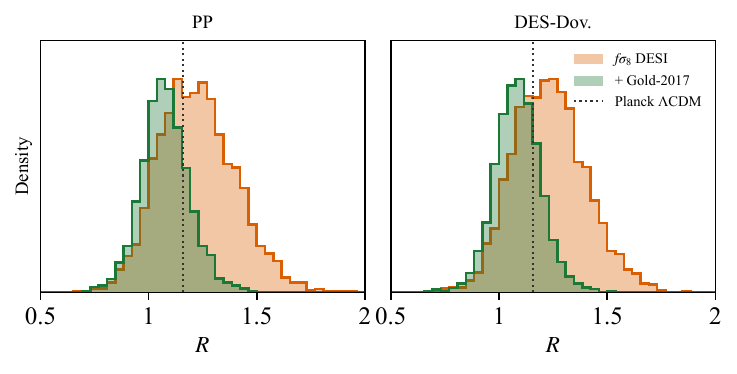}
    \caption{Posterior distribution of $R=(1+\zt)/(1+\zpk)$ for
    PP (left) and DES-Dov. (right) SNIa, with DESI DR1 $\fs$ only
    (orange) and with Gold-2017 added (green). The dotted vertical
    line marks $\RL=1.1596$ for Planck~2018 $\Lambda$CDM.}
    \label{S:fig:Rsamps}
\end{figure*}


\section{Reconstruction of the expansion and growth histories}
\label{S:sec:reconstruction}

We reconstruct $D_M(z)/r_d$, $D_H(z)/r_d$, and $\fs(z)$ directly
with a Multitask Gaussian process placed on the observables~\cite{S:Mukherjee:2024ryz,S:Mukherjee:2024pcg}. We use DR2 BAO distances including the Ly$\alpha$ AP+BAO
measurement~\cite{S:DESIDR2,S:DESIDR2LyaAP}, combined with DR1 $\fs$
measurements, and either Pantheon+~\cite{S:PantheonPlus} or
DES-Dovekie~\cite{S:DESDovekie} supernovae, optionally with the Gold-2017
RSD compilation~\cite{S:Gold2017}. 
The background task imposes spatial flatness,
\begin{equation}
\frac{\mathrm{d}}{\mathrm{d}z}\!\left(\frac{D_M}{r_d}\right)=\frac{D_H}{r_d},
\label{S:eq:flatness}
\end{equation}
through the derivative covariances of the distance GP; the growth
observable $\fs$ is a separate, independent task. Only when Gold RSD data are included, the AP effect couples the background to the growth sector through the angular diameter distance and expansion history; otherwise the two tasks are fully decoupled. 
Both tasks make use of squared-exponential kernels and
constant mean function with a template. In the matter-dominated limit, $f\to1$ and $\sigma_8\propto(1+z)^{-1}$,
so the growth GP is anchored by $\fs(z)\to A/(1+z)$ at high $z$,
where $A$ is a free amplitude. The GP hyperparameters, $A$,
$D_{H,0}/r_d$, and the supernova absolute magnitude are all
marginalized over in the joint posterior.
The supernova absolute magnitude is fully degenerate with $r_d$.
Hence, the sound horizon sets only the absolute distance scale.
The shapes of $q(z)$ and $\fs(z)$, and therefore $z_t$, $\zpk$,
and $R$, are independent of this choice. The joint posterior is sampled with the No-U-Turn Sampler
(NUTS)~\cite{S:Hoffman2014}, using two chains with 
1000 warm-up and 1500 retained samples each. For each posterior sample we locate the turnover in the expansion history, $z_t$, as the maximum of $\dfrac{(D_H/r_d)}{1+z}$, equivalently the zero of
\begin{equation}
q(z)=-1-(1+z)\frac{\mathrm{d}\ln(D_H/r_d)}{\mathrm{d}z},
\label{S:eq:q_recon}
\end{equation}
and the turnover in the growth history, $\zpk$, as the interior maximum of $\fs(z)$. The ratio $R=\dfrac{1+z_t}{1+\zpk}$ is finally computed directly from the reconstructed expansion and growth histories.

Figure~\ref{S:fig:recz} shows the reconstructed $q(z)$ and
$\fs(z)$.
In all cases $q(z)$ changes sign at $z\simeq0.7$, well above
the Planck-2018 $\Lambda$CDM value $z_t=0.632$.
The growth observable has a broad maximum at $z\simeq0.4$--$0.6$;
Gold-2017 sharpens the turnover (Fig.~\ref{S:fig:Rsamps}).
Table~II in the \textit{letter} lists the $68\%$ constraints.
The transition epoch is tightly constrained from BAO and SNIa, $z_t \simeq 0.72 \pm 0.07$. The growth turnover is less precise: $\zpk\simeq0.42$ excluding Gold-2017 and $\zpk\simeq0.60$ including Gold-2017, with uncertainties $\sim0.13$--$0.21$. Therefore, the error on $R$ is dominated by $\zpk$. The ratio is $R\simeq1.21$--$1.22$ without Gold and $R\simeq1.07$--$1.08$ with Gold-2017, with uncertainties $\sim0.17$ and $\sim0.10$ respectively; all estimates are consistent with $\RL=1.1596$ within $1\sigma$ CL.

%

\end{document}